\documentclass[pra,twocolumn,tightenlines,showpacs,nofootinbib]{revtex4}
\usepackage{bm,dcolumn,amsmath,graphicx}
\usepackage{epsfig}

\begin{document}
\title{Calculation of the (T,P)-odd Electric Dipole Moment of Thallium}

\author{V. A. Dzuba}
\affiliation{School of Physics, University of New South Wales,
                  Sydney, NSW 2052, Australia}
\author{V. V. Flambaum}
\affiliation{School of Physics, University of New South Wales,
                  Sydney, NSW 2052, Australia}

\date{\today}

\begin{abstract}

Parity and time invariance violating electric dipole moment of
$^{205}$Tl is calculated using the 
relativistic Hartree-Fock and configuration interaction methods and
the many-body perturbation theory. Contributions from the interaction
of the electron electric dipole moments with internal electric field
and scalar-pseudoscalar electron-nucleon (T,P)-odd interaction are
considered. The results are $d(^{205}{\rm Tl})=-582(20) d_e$ or
$d(^{205}{\rm Tl})=-7.0(2)\times 10^{-18}C^{SP} e \ {\rm cm}$.
Interpretation of the measurements are discussed.
The results of similar calculations for $^{133}$Cs are 
$d(^{133}{\rm Cs})=124(4) d_e$ or
$d(^{133}{\rm Cs})=0.76(2)\times 10^{-18}C^{SP} e \ {\rm cm}$.

\end{abstract}

\maketitle

\section{Introduction}

Recent very sensitive experiment performed in Seattle~\cite{Griffith}
puts very strong constrain on the electric dipole moment (EDM) of
mercury. It now reads $d(^{199}{\rm Hg}) = (0.49 \pm 1.29_{\rm stat} \pm 0.76_{\rm syst})
\times 10^{-29}\, e$ cm, which is sevenfold improvement of the
previous result of the same group. This renews the interest on the sources of
atomic EDMs.  In our previous paper~\cite{Dzuba09} we calculated the EDM of mercury
and other paramagnetic atoms due to nuclear Schiff moment,
(T,P)-odd electron-nucleon interaction and interaction of the electron
electric dipole moment ($d_e$) with nuclear magnetic field. The EDM of
mercury due to nuclear Schiff moment was also considered in a recent
paper by Latha {\em et al}~\cite{Latha}. Other contributions include,
e.g. interaction of the electron electric dipole moments with internal
electric field and scalar-pseudoscalar electron-nucleon (T,P)-odd
interaction. The latter two sources of atomic EDM are strongly
suppressed in mercury due to zero total electron momentum, $J=0$.
They give rise to EDMs of atoms with closed electron shells only in
third order of the perturbation theory, when magnetic dipole hyperfine
interaction is also taken into account. The strongest constrain on the
strength of these (T,P)-odd interactions came so far from the thallium
experiment~\cite{Regan} (see also review \cite{Ginges} for a detailed
discussion). However, significant advance in the accuracy of the
measurements in mercury \cite{Griffith} has changed the situation. Now
the constrain on the scalar-pseudoscalar electron-nucleon interaction
which comes from mercury EDM measurements is five times stronger than
those from thallium measurements while the constrains on the electron
EDM differ two times only~\cite{Griffith}. All these results rely on
atomic calculations which provide the link between atomic EDMs and the
fundamental constants of the (T,P)-odd interactions. Due to
significant progress in measurements it is important to revisit the
calculations as well for the sake of improving their accuracy and
reliability. 

The third-order calculations for mercury will be the subject of future
work. In present paper we perform calculations of the thallium EDM
caused by the electron EDM and the scalar-pseudoscalar
electron-nucleon (T,P)-odd interaction. The latter effect was
considered by M{\aa}rtensson-Pendrill and Lindroth \cite{Pendrill} and
Sahoo {\em et al}~\cite{Sahoo}. The results differ almost two times,
which is probably significantly larger than assumed uncertainty of
both calculations. 

The EDM of thallium caused by electron EDM was considered by many
authors \cite{Sandars,Flambaum,Johnson,Kraft,Hartley,Liu}. The results
show strong dependence on electron correlations and change
significantly depending on how many correlation terms are
included. The most complete calculations were performed by Liu and
Kelly \cite{Liu} using the coupled cluster approach. The
result is in relatively good agreement with the semiempirical
estimations of Ref.~\cite{Flambaum}.

All previous calculations of the thallium EDM treated the thallium
atom as a system with one external electron above closed shells. In
present paper we consider it as a three valence electron system by
including $6s$ electrons into valence space. We use the
configuration interaction technique combined with the many-body
perturbation theory (the CI+MBPT~\cite{Kozlov96,Dzuba96} method). We
demonstrate that all instabilities of the results are due to strong
correlations between external $6s$ and $6p$ electrons and using the
configuration interaction technique to treat these correlations
accurately leads to very stable results. We use exactly the same
procedure for both (T,P)-odd operators which is another test of the
consistency of the calculations. Our final result for the electron
EDM is in excellent agreement with the most complete previous {\em ab
  initio} calculations by Liu and Kelly~\cite{Liu}, while our result
for the scalar-pseudoscalar electron-nucleon (T,P)-odd interaction is
closer to the result of M{\aa}rtensson-Pendrill and
Lindroth~\cite{Pendrill} and differs significantly from Ref.~\cite{Sahoo}.

\section{Method of calculation}
\label{Sec:Method}

The Hamiltonian of the scalar-pseudoscalar electron-nucleon (T,P)-odd
interaction can be written as
\begin{equation}
  \hat H^{SP} = i\frac{G}{\sqrt{2}}AC^{SP}\gamma_0\gamma_5\rho_N(r),
\label{eq:s-p}
\end{equation}
where $G$ is the Fermi constant, $A=Z+N$ is nuclear mass number, 
$C^{SP}=(ZC^{SP}_p +NC^{SP}_n)/A$, and $\gamma$ are Dirac matrices.

The Hamiltonian for the electron EDM interacting with internal atomic
electric field $\mathbf{E}_{\rm int}$ can be written
as~\cite{Flambaum,Schiff,Sandars68}
\begin{equation}
  \hat H_e =
  -d_e\sum_{i=1}^Z(\gamma_0-1)^i\mathbf{\Sigma}^i\cdot\mathbf{E}^i_{\rm int}.
\label{eq:he2}
\end{equation}
Summation is over atomic electrons. 

Atomic EDM caused by any of the interactions
(\ref{eq:s-p},\ref{eq:he2}) is given by
\begin{equation}
\mathbf{d}_{\rm atom} = 2 \sum_M \frac{\langle 0
  |\mathbf{D}|M\rangle\langle M|\hat H^{TP}|0\rangle}{E_0-E_M},
\label{eq:datom}
\end{equation} 
where $|0\rangle$ is atomic ground state $\mathbf{D} =
-e\sum_i\mathbf{r}_i$ is the electric dipole operator and $\hat
H^{TP}$ is the (T,P)-odd operator. 

\subsection{The CI+MBPT method}

To calculate the EDM of thallium we consider it as a system with three
valence electrons above closed shells and use the CI+MBPT
method~\cite{Kozlov96,Dzuba96} for the valence electrons. The EDM of the atom
in the CI+MBPT is given by the formula very similar to
(\ref{eq:datom}) by with slightly different meaning of the
notations. First, the many electron states $|0\rangle$, $|M\rangle$
are now three-electron states in the valence space. Second, the
summation in the electric dipole operator $\mathbf{D}$ goes over
valence electrons only while contribution from atomic core is taken
into account by modifying the single-electron operator $\mathbf{d}$:
$\mathbf{d} \rightarrow \mathbf{d} + \delta V_{\rm core}$, where
$\delta V_{\rm core}$ is the correction to the electron core potential
caused by external field. Closed shell core does not contribute to the
EDM in the second order due to zero total angular momentum.

To perform the calculations we need to go through the following steps:
(a) generate a complete set of single-electron states; (b) build an
effective Hamiltonian in the valence space; (c) calculate core
polarization; (d) perform summation as in (\ref{eq:datom}) over a
complete set of three-electron states. Let's consider these tasks in
turn. 

We use the $\hat V^{N-3}$ approximation as in Ref.~\cite{Dzuba05}.
The calculations start from the relativistic
Hartree-Fock procedure for the triple ionized thallium ion. This gives
us the states and potential $\hat V_{\rm core} \equiv \hat V^{N-3}$ of the thallium core. 
We use the B-spline technique~\cite{B-spline} to generate a complete
set of single-electron states. These states are eigenstates of the
Dirac operator with the electron potential $\hat V^{N-3}$. We use
50 B-splines of order 9 in a cavity of radius 40$a_B$.

The effective CI+MBPT Hamiltonian for three valence electrons has the form
\begin{equation}
  \hat H^{\rm eff} = \sum_{i=1}^3 \hat h_1(r_i) + \sum_{i<j}^3
  \hat h_2(r_i,r_j), 
\label{Heff}
\end{equation}
where $\hat h_1$ is the single-electron part of the relativistic Hamiltonian
\begin{equation}
  \hat h_1 = c \mathbf{\hat{\alpha}} \mathbf{p} + (\hat{\beta}-1)m_e c^2-\frac{Ze^2}{r}
  + \hat V^{N-3} + \hat \Sigma_1,
\label{h1}
\end{equation}
and $\hat h_2$ is the two-electron part of the Hamiltonian
\begin{equation}
  \hat h_2(r_1,r_2) = \frac{e^2}{|\mathbf{r}_1 - \mathbf{r}_2|} + \hat
  \Sigma_2(r_1,r_2).
\label{h2}
\end{equation}
In these equations, $\mathbf{\hat{\alpha}}$ and $\hat{\beta}$ are the Dirac matrices,
$\hat V^{N-3}$ is the Dirac-Hartree-Fock (DHF) potential of the closed-shell atomic
core ($N-3=78,Z=81$), and $\hat \Sigma$ is the correlation operator. It
represents terms in the Hamiltonian arising due to virtual excitations from
atomic core (see Ref.~\cite{Kozlov96,Dzuba96} for details).
$\hat \Sigma \equiv 0$ corresponds to the standard CI method.
$\hat \Sigma_1$ is a single-electron operator. It represents a
correlation interaction of a particular valence electron with the atomic
core. $\hat \Sigma_2$ is a two-electron operator. It represents
screening of the Coulomb interaction between the two valence electrons by the core
electrons. We calculate  $\hat \Sigma_1$ for $s$-electrons using the
all-order technique developed in Ref.~\cite{Dzuba89}. $\hat \Sigma_1$
for $p$ and $d$ electrons as well as $\hat \Sigma_2$ are calculated 
in the second order of the many-body perturbation theory using the
B-spline basis set described above. We use 40 lowest B-spline states
up to $l_{\rm max}=5$ to calculate $\hat \Sigma$.

The same B-spline states are used to construct three-electron states
for valence electrons. We use 16 lowest states above the core up to 
$l_{\rm max}=2$ for this purpose. The basis for the ground state is
generated by allowing all possible single and double excitations from two initial
configurations $6s^26p$ and $6s6p6d$. The basis for even states
is generated by allowing all possible single and double excitations from three
initial configurations, $6s^27s$, $6s^26d$ and $6s6p^2$. Variation
of the basis size indicate that it is saturated with respect to 
$n_{\rm max}$ but not completely saturated with respect to 
$l_{\rm max}$. However, the contributions of the states with
$l_{\rm max}>2$ are small and can be neglected at required level of
accuracy. 

The three-electron valence states are found by solving the eigenvalue
problem,
\begin{equation}
  \hat H^{\rm eff} \Psi_v = E_v \Psi_v \, ,
\label{eq:CI}
\end{equation}
using the standard CI techniques. Calculated and experimental energies
of a few lowest-energy states of Tl are presented in Table~\ref{T:en}.
One can see that the inclusion of $\hat \Sigma$ (CI+MBPT) leads to
significant improvement of the agreement between theory and experiment.

\begin{table}
\caption{Three-electron removal energy (RE, a.u.) and excitation energies
  (cm$^{-1}$) of thallium.}
\label{T:en}

\begin{ruledtabular}
\begin{tabular}{lcrrr}
 \multicolumn{2}{c}{State} &\multicolumn{2}{c}{Theory} &
 Experiment\tablenotemark[1] \\
 & & \multicolumn{1}{c}{CI} & \multicolumn{1}{c}{CI+MBPT} & \\
\hline
\multicolumn{2}{l}{RE}    & -1.9177 & -2.0677 & -2.0722  \\
\hline
$6s^26p$ & $^2$P$^o_{1/2}$ &     0 &    0  &     0.0    \\ 
         & $^2$P$^o_{3/2}$ &  6345 &  8049  &   7793  \\ 
 	 	 	            
$6s^27s$ & $^2$S$_{1/2}$   & 23023 & 26810  &  26478  \\             
 	 	 	             
$6s^27p$ & $^2$P$^o_{1/2}$ & 30635 & 34496  &  34160  \\ 
         & $^2$P$^o_{3/2}$ & 31541 & 35507  &  35161  \\   
 	 	 	          
$6s^26d$ & $^2$D$_{3/2}$   & 32313 & 36553  &  36118  \\ 
         & $^2$D$_{5/2}$   & 32363 & 36624  &  36200  \\  
 	 	 	          
$6s^28s$ & $^2$S$_{1/2}$   & 34893 & 39037  &  38746  \\             
 	 	 	             
$6s^28p$ & $^2$P$^o_{1/2}$ & 37572 & 41714  &  41368  \\ 
         & $^2$P$^o_{3/2}$ & 37936 & 42122  &  41741  \\   
 	 	 	          
$6s^27d$ & $^2$D$_{3/2}$   & 38048 & 42359  &  42011  \\ 
         & $^2$D$_{5/2}$   & 38074 & 42395  &  42049  \\  
                                  
$6s^29s$ & $^2$S$_{1/2}$   & 39563 & 43728  &  43166  \\              
                                    
$6s^28d$ & $^2$D$_{3/2}$   & 40796 & 45931  &  44673  \\ 
         & $^2$D$_{5/2}$   & 41593 & 45971  &  44693  \\  
                                  
$6s6p^2$ & $^4$P$_{1/2}$   & 37195 & 43545  & 45220  \\ 
         & $^4$P$_{3/2}$   & 40797 & 48339  & 49800  \\  
         & $^4$P$_{5/2}$   & 44665 & 52779  & 53050  \\
                                  
\end{tabular}
\tablenotetext[1]{Reference \cite{Moore}}
\end{ruledtabular}
\end{table}
To calculate transition amplitudes we need to take into account the
effect of core polarization by external field. This is done by means
of the time-dependent relativistic Hartree-Fock method (see,
e.g. Ref.~\cite{Dzuba84}) which is equivalent to the random-phase
approximation, so we will use the term RPA for short.
The RPA equations for an external field operator $\hat F$ 
\begin{equation}
  (\hat h_1 -\epsilon_c)\delta \psi_c = -(\hat F + \delta \hat V^{N-3}_F)
  \psi_c
\label{Eq:RPA}
\end{equation}
are solved self-consistently for all states in atomic core in the
same $V^{N-3}$ potential as for the DHF states. The operator $\hat F$
is either the electric dipole operator or the operator of the
(T,P)-odd interaction, or any other operator (e.g. hyperfine
interaction). The correction to the core potential $\delta \hat V^{N-3}_F$
is used to calculate transition amplitudes 
\begin{equation}
  E1_{vw} = \langle \Psi_v | \hat F + \delta \hat V^{N-3}_F |
  \Psi_w \rangle. \label{E1} 
\end{equation}
Here $\Psi_v$ and $\Psi_w$ are three-electron states found by solving
the CI equations (\ref{eq:CI}). 

Calculated and experimental values of the electric dipole transition
amplitudes and magnetic dipole hyperfine structure (hfs) constants $A$ for
low states of thallium which are relevant to the calculation of the
EDM are presented in Table~\ref{T:e1}. Calculation of the hyperfine
structure is a good way to test the wave function on short distances
which is important for the matrix elements of weak interaction. The
data in the Table show that the accuracy of the calculation of the
E1-transition amplitudes and hyperfine constants of $s$ and $p_{1/2}$
states is within few percent. 

\begin{table}
\caption{E1 transition amplitudes and hfs constants $A$ of some low states
of $^{205}$Tl.}
\label{T:e1}
\begin{ruledtabular}
\begin{tabular}{lrrl}
\multicolumn{1}{c}{States} & \multicolumn{1}{c}{Calc.} &
 \multicolumn{2}{c}{Experiment} \\
\hline
\multicolumn{4}{c}{E1 transition amplitudes (a.u.) } \\
$6p_{1/2} - 7s_{1/2}$ & 1.73 & 1.81(2) & Ref.~\cite{Gallagher,Hsieh} \\
$6p_{1/2} - 6d_{3/2}$ & 2.23 & 2.30(9) & Ref.~\cite{Gallagher,Hsieh} \\
\multicolumn{4}{c}{Hyperfine structure constants $A$ (MHz) } \\ 
$6p_{1/2}$ & 21067 & 21311 & Ref.~\cite{Lurio} \\
$7s_{1/2}$ & 11417 & 12297 & Ref.~\cite{Odintzov} \\
\end{tabular}
\end{ruledtabular}
\end{table}

Finally, the last task we must be able to do to calculate the EDM is to
perform the summation over complete set of three-electron states. 
We use the Dalgarno-Lewis method~\cite{DalLew55} for this purpose
In this method, a correction $\delta \Psi_v$ to the three-electron wave
function of the ground state $v$ is introduced and the EDM
 is expressed as 
\begin{equation}
  \mathbf{d}_{\rm atom} = 2\langle \delta \Psi_v | \hat
  F + \delta \hat V^{N-3}_F | \Psi_v   \rangle \, . 
\label{eq:deltapsi}
\end{equation}
Here $\hat F$ is either the electric dipole operator or the operator of the
(T,P)-odd interaction. The correction $\delta \Psi_v$ is found by
solving the system of linear inhomogeneous equations
\begin{equation}
  (\hat H^{\rm eff} - E_v )\delta \Psi_v = - (\hat
  G+\delta \hat V^{N-3}_G) \Psi_v. 
\label{eq:DL}
\end{equation}
Here $\hat G$ is another operator from the pair $\mathbf{d}$, $H^{TP}$.
If both operators $\hat F$ and $\hat G$ are the same the electric dipole
operator $\mathbf{d}$, then the expression similar to
(\ref{eq:deltapsi}) gives static polarizability of the
atom. Table~\ref{T:pol} presents the results of the calculation of the
static scalar polarizability $\alpha_0$ of the thallium ground
state. Here $\alpha_{\rm core}$ is the contribution of the thallium
core to the polarizability, $\delta \alpha_{\rm core}$ is the
correction to the core polarizability due to Pauli principle which
forbids excitations from the core to the occupied $6s$ and $6p$
states, $\alpha _{\rm val}$ is the contribution of the valence
electrons to the polarizability. The final result is in good agreement
with other CI+MBPT~\cite{Kozlov01} and coupled cluster~\cite{Safronova06}
calculations. 

\begin{table}
\caption{Static scalar polarizability $\alpha_0$ of the thallium
  ground state (a.u.).}
\label{T:pol}
\begin{ruledtabular}
\begin{tabular}{ccccc}
$\alpha_{\rm core}$ & $\delta \alpha_{\rm core}$ & $\alpha_{\rm val}$
& Total & Other \\
\hline
4.98 & -0.67 & 44.50 & 48.81 & 49.2\tablenotemark[1],50.4\tablenotemark[2] \\
\end{tabular}
\tablenotetext[1]{Reference \cite{Kozlov01}}
\tablenotetext[1]{Reference \cite{Safronova06}}
\end{ruledtabular}
\end{table}

\section{Results}

The results of the calculations of the EDM of thallium in different
approximations are presented in Table~\ref{T:EDM} together with earlier
calculations. 
As it was pointed out in Ref.~\cite{Flambaum,Johnson,Kraft} thallium EDM 
is very sensitive to the strong correlations between $6s$ and $6p$ electrons. 
This interaction is treated pretty accurately in the configuration interaction 
technique used in present work. 
In contrast, all previous calculations treated
thallium as a system with one external electron above closed shells. 
Therefore, present results are significantly more stable than earlier
{\em ab initio} calculations. 

The main source of uncertainty for present calculations comes from the
core-valence correlations. Most of the core-valence correlations are
included via second-order correlation operator $\hat \Sigma$.
However, there are small contributions like higher-order correlations,
correction to $\hat \Sigma$ due to external field (structure
radiation), renormalization of the wave function, etc. Quantum
electrodynamic and Breit corrections are also expected to be
small~\cite{Dzuba06}. As one can see from Table~\ref{T:EDM} the effect
of including $\hat \Sigma$ into full-scale CI calculations on the EDM
of Tl is about 3\%. We use this as an estimate of the accuracy of our calculations.

Similar calculations for cesium give the following results (in
agreement with previous calculations, see review~\cite{Ginges}):
\begin{equation}
 d({\rm Cs}) = 0.759 \times 10^{-18} C^{SP} e \ {\rm cm}, 
\end{equation}
or
\begin{equation}
 d(\rm Cs) = 124 \ d_e.
\end{equation} 
The estimated error of these results is about 3\%:

\begin{table}
\caption{EDM of Tl due to electron EDM ($d_e$) and scalar-pseudoscalar
  electron-nucleon (T,P)-odd interaction.} 
\label{T:EDM}
\begin{ruledtabular}
\begin{tabular}{lll}
\multicolumn{1}{c}{$d_e$} & \multicolumn{1}{c}{$10^{-18}C^{SP} e$ cm} & Comments\\
\hline
\multicolumn{3}{c}{this work}  \\
-614 & -7.33 & single-configuration, no $\hat \Sigma$ \\ 
-537 & -6.43 & single-configuration with $\hat \Sigma$ \\ 
-625 & -7.49 & single-configuration \\
     &      & in the ground state, with $\hat \Sigma$ \\ 
-602 & -7.22 & full CI but no $\hat \Sigma$ \\
-581 & -6.88 & full CI+MBPT but no RPA \\
-582 & -6.98 & full scale calculations \\
-582(20) & -7.0(2) & final \\
\multicolumn{3}{c}{other calculations} \\
-585 &      & Ref.~\cite{Liu}, coupled cluster \\
-1041&      & Ref.~\cite{Johnson}, DHF+1st order MBPT \\
-502 &      & Ref.~\cite{Johnson}, Tietz\tablenotemark[1]+1st order MBPT \\
-607 &      & Ref.~\cite{Johnson}, Green\tablenotemark[1]+1st order MBPT \\
-562 &      & Ref.~\cite{Johnson}, Norcross\tablenotemark[1]+1st order MBPT \\
 700 &      & Ref.~\cite{Sandars}, parametric potential \\
-500 &      & Ref.~\cite{Flambaum}, semiempirical estimate \\
-301 &      & Ref.~\cite{Kraft},  2nd order  MBPT         \\
-179 &      & Ref.~\cite{Hartley},  2nd order  MBPT           \\
     & -4.056 & Ref.~\cite{Sahoo}, coupled cluster \\
     & -7(2) & Ref.~\cite{Pendrill}, RPA+rescaling       \\
     &      & of correlations from Ref.~\cite{Hartley} \\
\end{tabular}
\tablenotetext[1]{Parametric potentials}
\end{ruledtabular}
\end{table}
The result of measurement of the EDM of $^{205}$Tl~\cite{Regan} reads
\begin{equation}
  d(^{205}{\rm Tl}) = -(4.0 \pm 4.3) \times 10^{-25} e \ {\rm cm} .
\label{eq:tledm}
\end{equation}
Using the numbers from Table~\ref{T:EDM} we find 
\begin{equation}
  d_e = (6.9 \pm 7.4) \times 10^{-28} e \ {\rm cm},
\label{eq:de}
\end{equation}
and
\begin{equation}
  C^{SP} = (5.7 \pm 6.2) \times 10^{-8} .
\label{eq:csp}
\end{equation}
These numbers are in good agreement with the analysis of Ref.~\cite{Ginges}.

\section*{Acknowledgment}

The work was supported in part by the Australian Research Council.


\begin{thebibliography}{99}
\bibitem{Griffith} W. C. Griffith, M. D. Swallows, T. H. Loftus,
 M. V. Romalis, B. R. Heckel, and E. N. Fortson, 
Phys. Rev. Lett. {\bf 102}, 101601 (2009).

\bibitem{Dzuba09} V. A. Dzuba, V. V. Flambaum, and S. G. Porsev,
Phys. Rev. A {\bf 80}, ??????   (2009).
	
\bibitem{Latha} K. V. P. Latha, D. Angom, B. P. Das, and D. Mukherjee,
Phys. Rev. Lett. {\bf 103}, 083001 (2009).

\bibitem{Regan} B. C. Regan, E. D. Commins, C. J. Schmidt, D. DeMille,
Phys. Rev. Lett. {\bf 88}, 071805 (2002). 

\bibitem{Ginges} J. S. M. Ginges and V. V. Flambaum, Phys. Rep. {\bf
    397}, 63 (2004).

\bibitem{Sandars} P. G. H. Sandars and R. M. Sternheimer,
Phys. Rev. A {\bf 11}, 473 (1975).

\bibitem{Flambaum} V. V. Flambaum, Yad. Fiz. {\bf 24}, 383 (1976)
  [Sov. J. Nuc. Phys. {\bf 24}, 199 (1976)].

\bibitem{Johnson} W. R. Johnson, D. S. Guo, M. Idrees, and
  J. Sapirstein, Phys. Rev. A {\bf 34}, 1043 (1986).

\bibitem{Kraft} A. Ya. Kraftmakher, J. Phys. B {\bf 21}, 2803 (1988).

\bibitem{Hartley} A. C. Hartley, E. Lindroth, and
  A.-M. M{\aa}rtensson-Pendrill,  J. Phys. B {\bf 23}, 3417 (1990).

\bibitem{Liu} Z. W. Liu and H. P. Kelly, Phys. Rev. A {\bf 45}, R4210 (1992). 

\bibitem{Sahoo} B. K. Sahoo, B. P. Das, R. K. Chaudhuri, D. Mukherjee,
  and E. P. Venugopal, Phys. Rev. A {\bf 78}, 010501(R) (2008); {\bf
    78}, 039901(E) (2008).

\bibitem{Pendrill}  A.-M. M{\aa}rtensson-Pendrill and E. Lindroth,
  Europhys. Lett. {\bf 15}, 155 (1991).

\bibitem{Kozlov96} V. A. Dzuba, V. V. Flambaum, and M. G. Kozlov,
Phys. Rev. A, {\bf 54}, 3948 (1996).

\bibitem{Dzuba96} V. A. Dzuba, V. V. Flambaum, and M. G. Kozlov,
JETP Lett., {\bf 63}, 882 (1996).

\bibitem{Schiff} L. I. Schiff, Phys. Rev. {\bf 132}, 2194 (1963).

\bibitem{Sandars68} P. G. H. Sandars, J. Phys. {\bf 1B}, 511 (1968). 

\bibitem{Dzuba05} V. A. Dzuba, 
Phys. Rev. A, {\bf 71}, 032512 (2005).

\bibitem{B-spline} W. R. Johnson, and J. Sapirstein, Phys. Rev.
Lett. {\bf 57}, 1126 (1986).

\bibitem{Dzuba89} V. A. Dzuba, V. V. Flambaum, O. P. Sushkov,
Phys. Lett A., {\bf 140}, 493-497 (1989).

\bibitem{Dzuba84} V. A. Dzuba, V. V. Flambaum, O. P. Sushkov, 
J. Phys. B {\bf 17}, (1984).

\bibitem{Moore}   C. E. Moore,
        {\it Atomic Energy Levels} Natl. Bur. Stand. (US),
        Circ. No. 467 (U.S. GPO, Washington, DC, 1958), Vol. III.

\bibitem{Gallagher} A. Gallagher and A. Lurio, Phys. Rev. {\bf 136}, A87 (1964).

\bibitem{Hsieh} J. C. Hsieh and J. C. Baird, Phys. Rev. A {\bf 6}, 141 (1972).

\bibitem{Lurio} A. Lurio and A. G. Prodell, Phys. Rev. {\bf 101}, 79
  (1956).

\bibitem{Odintzov} A. I. Odintzov, Opt. Stectrosc. {\bf 9}, 142 (1960).

\bibitem{DalLew55} A. Dalgarno and J. T. Lewis,  Proc. R. Soc. London
  {\bf 233}, 70 (1955).

\bibitem{Kozlov01} M. G. Kozlov, S. G. Porsev, and W. R. Johnson, 
Phys. Rev. A {\bf 64}, 052107 (2001). 

\bibitem{Safronova06} M. S. Safronova, W. R. Johnson, U. I. Safronova, and
T. E. Cowan, Phys. Rev. A {\bf 74}, 022504 (2006). 

\bibitem{Dzuba06} V. A. Dzuba, V. V. Flambaum, M. S. Safronova,
      Phys. Rev. A, {\bf 73} 022112 (2006).

\end{thebibliography}
\end{document}